\documentclass[reprint,amsmath,amssymb,pra,aps, nofootinbib,superscriptaddress]{revtex4-1}

\usepackage[T1]{fontenc}
\usepackage[utf8]{inputenc}
\usepackage{amsmath, amssymb}
\usepackage{bbold}
\usepackage{bm}
\usepackage{tipa}
\usepackage{blindtext}
\usepackage[dvipsnames]{xcolor}
\definecolor{dcyan}{RGB}{0,100,100}
\usepackage{graphicx}
\usepackage{tabularx}
\usepackage{textgreek}
\usepackage{xfrac}
\usepackage{units}
\usepackage{comment}
\usepackage{float}

\usepackage[sectionbib]{bibunits}
\defaultbibliographystyle{naturemag}
\defaultbibliography{references}

\usepackage{hyperref}
\usepackage{xcolor}
\definecolor{green_cust}{RGB}{0,154,85}
\definecolor{red_cust}{RGB}{173,49,54}
\definecolor{blue_cust}{RGB}{0,103,148}
\hypersetup{colorlinks=true, linkcolor=red_cust, citecolor=green_cust, filecolor=blue_cust, urlcolor=blue_cust}

\usepackage{soul}

\newcommand{\ket}[1]{|{#1}\rangle}

\newcommand{\Figref}[1]{Fig.~\hyperref[#1]{\ref{#1}}}

\begin{document}
\title{Manybody Interferometry of Quantum Fluids}

\author{Gabrielle Roberts$^*$}
\author{Andrei Vrajitoarea$^*$}
\author{Brendan Saxberg}
\author{Margaret G. Panetta}
\affiliation{The Department of Physics, The James Franck Institute, and The Pritzker School of Molecular Engineering, The University of Chicago, Chicago, IL}
\author{Jonathan Simon}
\affiliation{The Department of Physics, The James Franck Institute, and The Pritzker School of Molecular Engineering, The University of Chicago, Chicago, IL}
\affiliation{The Department of Physics, Stanford University, Stanford, CA}
\affiliation{The Department of Applied Physics, Stanford University, Stanford, CA}
\author{David I. Schuster}
\affiliation{The Department of Physics, The James Franck Institute, and The Pritzker School of Molecular Engineering, The University of Chicago, Chicago, IL}
\affiliation{The Department of Applied Physics, Stanford University, Stanford, CA}
\date{\today}


\begin{abstract}
Characterizing strongly correlated matter is an increasingly central challenge in quantum science, where structure is often obscured by massive entanglement. From semiconductor heterostructures~\cite{manfra2014molecular} and 2D materials~\cite{cao2018unconventional} to synthetic atomic~\cite{gross2017quantum}, photonic~\cite{carusotto2020photonic,clark2020observation} and ionic~\cite{pagano2020quantum} quantum matter, progress in \emph{preparation} of manybody quantum states is accelerating, opening the door to new approaches to state characterization. It is becoming increasingly clear that in the quantum regime, state preparation and characterization should not be treated separately -- entangling the two processes provides a quantum advantage in information extraction. From Loschmidt echo~\cite{Braumuller_MITOTOC} to measure the effect of a perturbation, to out-of-time-order-correlators (OTOCs) to characterize scrambling~\cite{swingle2016measuring,landsman2019verified,googlescramble2021} and manybody localization~\cite{fan2017out}, to impurity interferometry to measure topological invariants~\cite{grusdt2016interferometric}, and even quantum Fourier transform-enhanced sensing~\cite{vorobyov2021quantum}, protocols that blur the distinction between state preparation and characterization are becoming prevalent. Here we present a new approach which we term ``manybody Ramsey interferometry'' that combines adiabatic state preparation and Ramsey spectroscopy: leveraging our recently-developed one-to-one mapping between computational-basis states and manybody eigenstates~\cite{AdbPaper}, we prepare a superposition of manybody eigenstates controlled by the state of an ancilla qubit, allow the superposition to evolve relative phase, and then reverse the preparation protocol to disentangle the ancilla while localizing phase information back into it. Ancilla tomography then extracts information about the manybody eigenstates, the associated excitation spectrum, and thermodynamic observables. This work opens new avenues for characterizing manybody states, paving the way for quantum computers to efficiently probe quantum matter.

\end{abstract}

\maketitle

\section{Introduction}
\label{sec:intro}
Advances in controllable quantum science platforms have opened the possibility of creating synthetic quantum materials, in which the physical laws governing the material are built-to-order in the lab~\cite{carusotto2020photonic, blatt2012quantum, bloch2012quantum, clark2020observation}. Such experiments enable time- and space-resolved probes~\cite{bakr2009quantum,cheuk2015quantum,browaeys2020many} of quantum dynamics inaccessible in solid-state matter, as well as explorations of extreme parameter regimes~\cite{karamlou2022quantum,wintersperger2020realization,barbiero2019coupling,kollar2019hyperbolic,greiner2002quantum}. Indeed, as the community has become increasingly adept at leveraging the flexibility of synthetic matter platforms to realize arbitrary physical laws, we now face the challenge of capitalizing on this same flexibility for preparing and characterizing quantum manybody states.

In electronic materials, preparing low-entropy equilibrium states relies upon \emph{refrigeration}: harnessing the coupling of the material to a low-temperature reservoir that can absorb its entropy. By contrast, synthetic material platforms are known for their coherent, low-dissipation evolution, and hence their lack of reservoir coupling. State preparation has thus relied upon the development of new approaches based upon engineered reservoirs~\cite{wineland1979laser,ketterle1996evaporative,poyatos1996quantum,barreiro2011open,Ma2019AuthorPhotons,carusotto2020photonic} and adiabatic evolution~\cite {greiner2002quantum,simon2011quantum,leonard2023realization}, elucidating, among other things, microscopic aspects of quantum thermodynamics~\cite{Hafezi2015ChemicalCoupling,kurilovich2022stabilizing} and the importance of symmetry breaking~\cite{he2017realizing}, respectively.

\begin{figure*}[ht]
	\centering
	\includegraphics[width=1\textwidth]{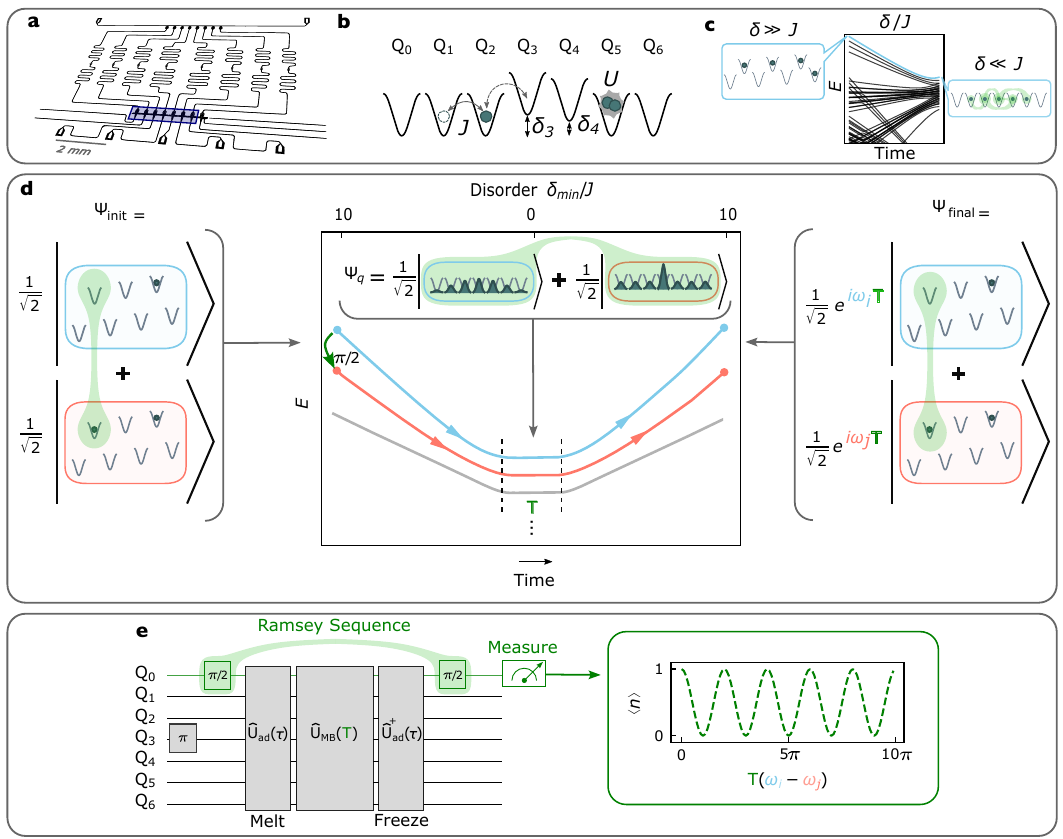}
	\caption{\textbf{Preparing and Interfering Manybody States}. The quantum system probed in this work consists of a chain of seven capacitively coupled superconducting transmon qubits~\cite{Koch2007} (blue in \textbf{a}) connected to site-resolved readout resonators (meandering traces) and flux control (bottom traces). \textbf{b,} The system is well described by the Bose-Hubbard model: particles (microwave photons) coherently tunnel between lattice sites (qubits) at a rate $J$, with on-site interactions $U\gg J$ arising from the transmon anharmonicity~\cite{Ma2019AuthorPhotons, carusotto2020photonic, Houck2012On-chipCircuits}. Real-time flux tuning provides control of lattice site energies ($\delta_i$ for site $i$), allowing the deterministic manipulation of disorder that we leverage to build highly entangled states: \textbf{c,} Starting in a highly disordered lattice, we initialize the system in the highest energy $N$-particle eigenstate by applying $\pi$-pulses to the $N$ highest-energy sites (left), and adiabatically removing disorder to convert these states into eigenstates of the quantum fluid (right)~\cite{AdbPaper}. \textbf{d,} To interfere superpositions of such states, we replace one of the assembly $\pi$-pulses with a $\frac{\pi}{2}$ pulse (green), producing a superposition of two (red/blue) eigenstates; adiabatically removing disorder produces a superposition of two manybody fluid states; coherently evolving for a time $T$ allows the eigenstates to accumulate a relative phase proportional to their energy difference; ramping back to the disordered configuration (right) relocalizes the phase difference into the single qubit that started in a superposition; a final $\frac{\pi}{2}$ pulse on this qubit maps the phase information onto qubit occupancy for measurement. \textbf{e} reinterprets the full manybody Ramsey sequence as a set of gates on the qubits comprising the lattice, resulting in an interference fringe vs evolution time $T$.}
    \label{fig:setupfig}
\end{figure*}

Even once a manybody state is prepared, characterizing it presents unique challenges. The intuitively simplest but technically most demanding characterization approach is state tomography, where n-body correlations are measured in complementary bases, allowing complete reconstruction of the system density matrix~\cite{roos2004control}. This approach has the advantage that \emph{all} information about the state is extracted, and the disadvantage that the required statistics (and thus measurement time) scale exponentially with system size. If \emph{specific} rather than \emph{complete} information about the state is desired, more carefully crafted protocols have been shown to relax measurement requirements: 
Expansion imaging measures single-particle coherence~\cite{ketterle1999making}; noise correlations measure two-body ordering~\cite{greiner2005probing}; in-situ density probes equation of state~\cite{Yefsah_Thermo2Dgas}; Bragg spectroscopy is sensitive to density- (and \hbox{spin-)} waves~\cite{ernst2010probing}; particle-resolved readout accesses higher-order correlations ~\cite{bakr2009quantum,hilker2017revealing, ibm2023integrability, karamlou2022quantum, caltech_integrability}; parity oscillations are clear signatures of GHZ states~\cite{monz201114}; single-qubit tomography probes global entanglement~\cite{karamlou2022quantum,AdbPaper};
and shadow tomography~\cite{huang2020predicting} provides an efficient way to extract observables from few measurements.

It has nonetheless become apparent that treating state \emph{preparation} and state \emph{characterization} as independent does not fully leverage quantum advantage -- approaches that entangle the two tasks via an ancilla can be vastly more performant: Proposals \& experiments to quantify scrambling~\cite{swingle2016measuring,landsman2019verified,googlescramble2021} and verify manybody localization~\cite{fan2017out} rely upon out-of-time-order correlators (OTOCs) that compare manybody states to which a specific operator is applied either before or after coherent evolution. This is achieved by entangling the time at which the operator is applied with the state of an ancilla, and subsequently performing tomography on the ancilla. Similarly, Loschmidt echoes directly measure the impact of perturbations via state overlap measurements following evolution under two similar Hamiltonians~\cite{Braumuller_MITOTOC,Xu2020phasetrans}. Sensitivity \& dynamic range enhancements in sensing~\cite{vorobyov2021quantum} can be achieved by sandwiching ancilla-conditioned dynamics between quantum Fourier transforms. Entangling initial states with an ancilla and applying ancilla-conditioned evolution can further probe anyon braiding phase~\cite{GoogleAnyons2021} and system spectrum~\cite{Morvan2022-boundstate, roushan2017spectroscopic}.

In this work we introduce \emph{manybody Ramsey interferometry} as a direct probe of thermodynamic observables: we entangle which manybody state we prepare in a Bose-Hubbard circuit with the state of an ancilla qubit, allow the superposition to evolve, disentangle from the ancilla, and perform ancilla tomography to learn about the manybody states. We rely upon our recently-demonstrated reversible one-to-one mapping of computational states onto manybody states~\cite{AdbPaper} to achieve the ancilla/manybody state entanglement. Because we entangle and then disentangle the ancilla from the manybody system, we localize the sought-after information in a single qubit for efficient, high signal-to-noise readout, rather than extracting it from a many-qubit state space~\cite{roushan2017spectroscopic,Morvan2022-boundstate, MBRamsey_PRL2013,zeiher2016many}.

In Section~\ref{sec:circuit&protocol} we introduce our circuit platform and manybody Ramsey protocol. In Section~\ref{sec:superpositions} we demonstrate the protocol and in Section~\ref{sec:adb} we use manybody Ramsey to probe adiabaticity of state preparation. Finally in Section~\ref{sec:thermo} we employ manybody Ramsey to directly measure thermodynamic observables of a strongly interacting quantum fluid by studying superpositions of (i) particle number and (ii) system size.

\section{The Platform}
\label{sec:circuit&protocol}

The properties of our synthetic quantum material platform are accurately captured by a 1D Bose-Hubbard model (see Fig.~\ref{fig:setupfig}b), describing bosonic particles tunneling between lattice sites at rate $J$, in the presence of onsite interactions of energy $U$,
\begin{align*}
\mathbf{H}_\mathrm{BH}(t)/\hbar & = J \sum_{ \langle i,j \rangle}{a_i^\dagger a_j } + \frac{U}{2}\sum_i{n_i \left(n_i-1\right)}\\
&+ \sum_i {(\omega_\mathrm{lat}+\delta_i(t)) n_i} \label{eq:bosehubbardC}.
\end{align*}

Our Hubbard lattice is realized in a quantum circuit~\cite{Ma2019AuthorPhotons, carusotto2020photonic, Houck2012On-chipCircuits}: sites are implemented as transmon qubits, particles as microwave photon excitations of the qubits, tunneling ($J$) as capacitive coupling between the qubits (Fig.~\ref{fig:setupfig}a,b), and onsite interactions ($U$) as transmon anharmonicity. Lattice site energies (qubit frequencies) can be individually \& dynamically tuned using flux bias lines (see SI~\ref{SI:DeviceFabandParams}). For this work $J/2\pi=-9$~MHz, $U/2\pi=-240$~MHz, and $\omega_{lat}/2\pi\approx 5$~GHz. The tuning range of our qubits extends from $\omega_{qb}/2\pi\sim 3-6$~GHz. The photon lifetime $T_1\approx 40$~\textmu s is much longer than the timescale of the manybody dynamics (see SI~\ref{SI:DeviceFabandParams} for details).

We recently demonstrated adiabatic preparation of photonic fluids by leveraging real-time ($\ll$ tunneling time) control of lattice disorder~\cite{AdbPaper}. This protocol begins with lattice sites tuned apart in energy by more than the tunneling $J$. In this configuration, the many-particle eigenstates are localized into product states over individual sites, such that any eigenstate may be prepared via site-resolved microwave $\pi$-pulses that inject individual photons. By next adiabatically removing the lattice disorder, we smoothly convert the localized eigenstates of the disordered system into the corresponding highly entangled eigenstates of the disorder-free system (Fig.~\ref{fig:setupfig}c). The combination of the one-to-one mapping and the ease of state preparation in the disordered (staggered) system render it straightforward to prepare \emph{any} eigenstate of the ordered system provided sufficient coherence time to ensure adiabaticity in the disorder-removal ramp.

We now harness this precise eigenstate preparation to explore controlled interference of many-particle quantum states. Our approach can be understood in analogy to traditional Ramsey spectroscopy of a single qubit~\cite{ramsey1990experiments} with states $\ket{0}$ and $\ket{1}$: in this simpler case, a system prepared in $\ket{0}$ is driven into an equal superposition of $\ket{0}$ and $\ket{1}$ with a $\frac{\pi}{2}$ pulse, and after an evolution time $T$, the phase accrued on between $\ket{0}$ and $\ket{1}$ is read out with a second, phase-coherent $\frac{\pi}{2}$ pulse: the resulting population difference between $\ket{0}$ an $\ket{1}$ states oscillates (versus evolution time $T$) at a frequency set by the energy difference between the states.

To extend the protocol to measurement of the phase difference between two manybody states, we take the single-qubit Ramsey sequence above and sandwich the delay time $T$ between qubit-conditioned assembly and disassembly of the manybody states. We call this approach 'manybody Ramsey interferometry' to connect with previous work exploring non-adiabatic evolution of manybody systems~\cite{googlescramble2021, MBRamsey_PRL2013,zeiher2016many}. This procedure maps the phase accrued between the manybody states \emph{entirely} onto the single qubit, avoiding any reduction in contrast due to residual entanglement, at measurement time, with the manybody system. The enabling ingredient for this protocol is qubit-conditioned manybody state preparation, which we implement via our disorder-assisted adiabatic assembly techniques~\cite{AdbPaper}.

An example of the full protocol is shown in Figure~\ref{fig:setupfig}d. In the presence of disorder, we prepare a superposition of the highest energy states of the $N=1$ and $N=2$ particle manifolds $\ket{\Psi_i}=\frac{1}{\sqrt{2}}(\ket{0000010}+\ket{0001010})$=$\ket{000} \otimes \frac{\ket{0}+\ket{1}}{\sqrt2} \otimes \ket{010}$. The key to this protocol is that in the presence of disorder the superposition of the two highest-energy manybody states is realized as a superposition of a \emph{single control qubit}, realized with a $\frac{\pi}{2}$ pulse. As we adiabatically remove the disorder the localized states melt into corresponding eigenstates of the quantum fluid. During the subsequent hold time $T$, these states will accrue a relative phase proportional to their energy difference, $(\omega_i - \omega_j)T$. Finally, to relocalize the information back into the control qubit we adiabatically re-introduce lattice disorder, producing the final state: $\ket{\Psi_f}=\frac{1}{\sqrt{2}}(\ket{0000010}+e^{i(\omega_i - \omega_j)T}\ket{0001010})=\ket{000} \otimes \frac{\ket{0}+e^{i(\omega_i - \omega_j)T}\ket{1}}{\sqrt2} \otimes \ket{010}$. The phase accrued between the manybody states has now been written entirely into the control qubit. We extract that phase information (and thus the manybody energy-difference) with a final $\frac{\pi}{2}$ pulse on the control qubit and a population measurement in the $\ket{0}$,$\ket{1}$ basis. The complete pulse sequence is illustrated in Fig.~\ref{fig:setupfig}e.

\begin{figure*}[ht] 
	\centering
 	\includegraphics[width=0.95\textwidth]{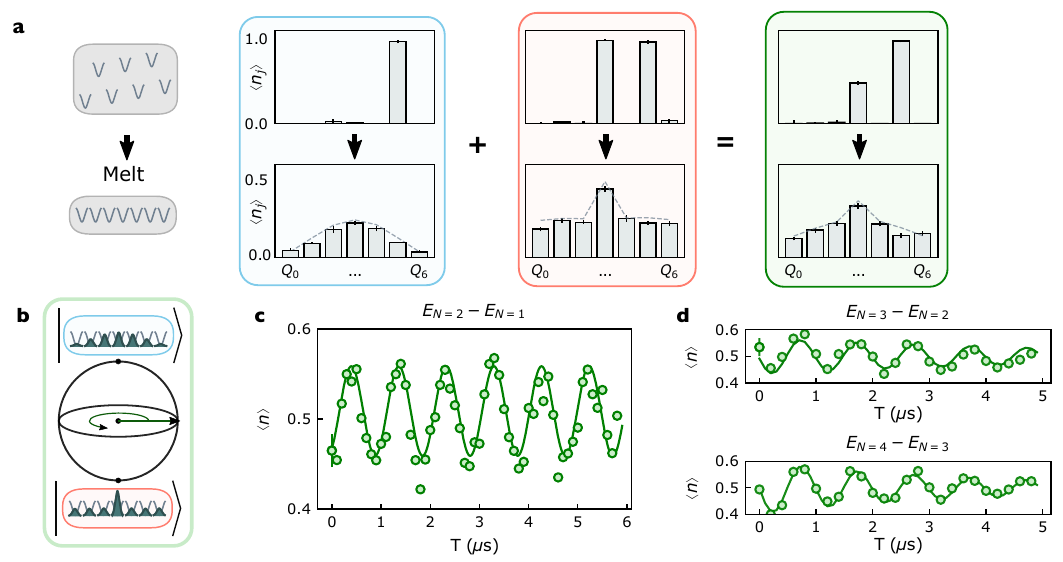}
	\caption{
		\textbf{Benchmarking the Manybody Ramsey Protocol}. \textbf{a,} To explore this protocol in the lab, we deterministically inject particles into a disordered lattice \& remove the disorder (left), before imaging the resulting density distribution. When we inject precisely 1 photon (upper blue panel), adiabatic disorder removal produces the lowest-momentum particle-in-a-box state (lower blue panel); injecting two photons (upper red panel) produces the lowest energy two-body state after disorder is removed (lower red panel). If we deterministically inject the first particle with a $\pi$ pulse but $\frac{\pi}{2}$ pulse the second photon, we should produce the manybody superposition state, and indeed we observe the average of the two density distributions (green panels). \textbf{b,} To demonstrate that this average density distribution corresponds to the macroscopic superposition of manybody states, we allow the superposition state to evolve on the Bloch sphere before adiabatically mapping the manybody superposition back onto a single qubit, where it can be read out via a second $\frac{\pi}{2}$ pulse. \textbf{c,} The resulting Ramsey fringe (vs hold time $T$ in the manybody superposition state) evolves with a frequency given by the energy difference between the manybody states minus the frequency of the local oscillator from which the $\frac{\pi}{2}$ pulses are derived, exhibiting contrast over several microseconds limited by the single-qubit $T_2$ (see SI~\ref{SI:DeviceFabandParams}). In \textbf{d,} we demonstrate the applicability of the approach to larger systems by applying it to superpositions of $N=2,3$ and $N=3,4$ particle fluids; the increased decay reflects the faster dephasing of states with more particles. Representative error bars (on first data point of each plot) reflect the S.E.M.
	}
	\label{fig:PrepSuperpositions}
\end{figure*} 

\section{Demonstration of the Protocol}
\label{sec:superpositions}

We benchmark our manybody Ramsey protocol by studying the superposition of $1$- and $2$- photon fluid ground states in our Hubbard circuit. In Fig.~\ref{fig:PrepSuperpositions}a we prepare these states separately (red and blue boxes) by $\pi$-pulsing localized particles into the disordered lattice \& then adiabatically removing the disorder, finding good agreement of the measured \emph{in-situ} density profiles with a parameter-free Tonks gas model~\cite{Kinoshita2004ObservationGas,Paredes2004TonksGirardeauLattice} (see SI ~\ref{SI:TGgas}). When the second particle is instead injected with a $\frac{\pi}{2}$ pulse, we create the desired superposition state, with a density profile reflecting the average of the two participating eigenstates (green box).

To measure the energy difference between these states we must interfere them. We achieve this by replacing the in-situ density measurement with a coherent evolution time $T$, allowing the states to accrue a relative phase (Fig.~\ref{fig:PrepSuperpositions}b), followed by adiabatically reintroducing the disorder to re-localize the phase information into a single lattice site (qubit) and finally applying a $\frac{\pi}{2}$ pulse to interfere the states \& read out the encoded phase in the occupation basis. The resulting sinusoidal Ramsey fringe (occupation vs $T$) is shown in Fig.~\ref{fig:PrepSuperpositions}c, with contrast limited by qubit dephasing (see SI~\ref{SI:DeviceFabandParams}). The fringe frequency of 10~MHz is translated down (for clarity) from the actual energy difference of 5.317~GHz via a $T$-dependent phase offset of the second $\frac{\pi}{2}$ pulse (see SI~\ref{SI:RamseyMeas}). Similar experiments enable single-qubit measurement of energy differences of 2/3 \& 3/4 particle superposition states (Fig.~\ref{fig:PrepSuperpositions}d), with minimal contrast reduction, and only a small drop in coherence time. We are thus prepared to apply the protocol to exploration of manybody physics.


\section{Probing the Excitation Spectrum}
\label{sec:adb}
\begin{figure*} 
	\centering
	\includegraphics[width=0.95\textwidth]{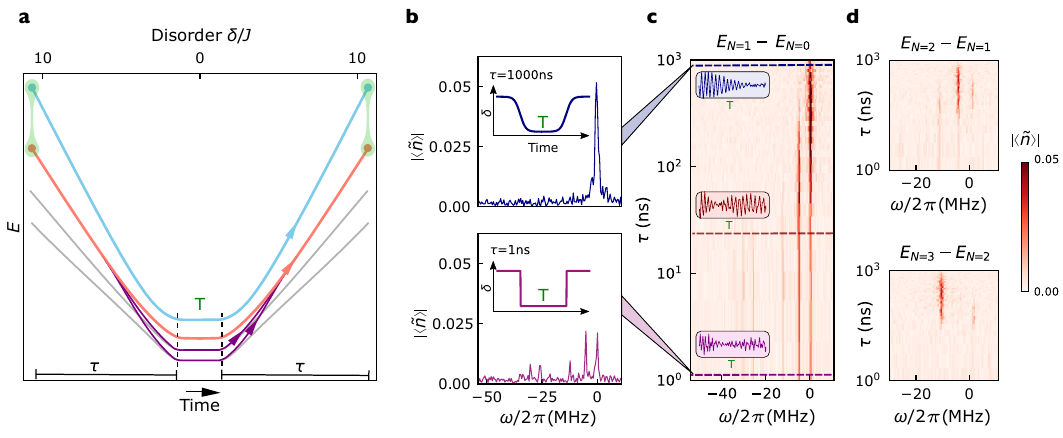}
	\caption[width=\textwidth]{
		\textbf{Spectroscopic Signatures of Adiabaticity}. The manybody Ramsey protocol relies critically on the ability to adiabatically map localized states into and out of highly entangled states (in a ramp time $\tau$). \textbf{a,} Ramping too quickly leads to diabatic excitations (purple) into other manybody states that do not interfere with the states (red/blue) in the prepared superposition (green) and thus reduce Ramsey fringe contrast. With some probability, however, these diabatic excitations are diabatically de-excited back into the initial state (red) during the backwards ramp; because these excitations evolve at different frequencies (corresponding to their energies) during the hold time $T$, they produce Ramsey fringes at other Fourier frequencies. \textbf{b,} For the slowest ramp ($\tau=1$~\textmu s) there are no diabatic excitations, producing a single Fourier feature in the Ramsey interference between $N=0$ and $N=1$ eigenstates. \textbf{c} For the fastest ramp ($\tau=1$~ns), the many diabatic excitations are reflected in additional frequencies in the Ramsey fringe beyond the dominant feature in the slow ramp. \textbf{c,} As the ramp time $\tau$ is varied over three decades, frequency components furthest from the dominant feature disappear first, with the low-offset-frequency features disappearing only for the slowest ramps, consistent with excitation rates controlled by the energy gaps of the fluid. \textbf{Insets} depict the time-domain Ramsey fringes for slow, intermediate, and fast ramps (top to bottom). \textbf{d,} Repeating these experiments with superpositions of $N=1,2$ and $N=2,3$ particles demonstrates that while the proliferation of manybody states makes resolving diabatic excitations challenging, the dominant feature nonetheless appears for the slowest ramps.
  }\label{fig:adb}
\end{figure*}

Our manybody Ramsey protocol relies upon our ability to adiabatically assemble and disassemble highly entangled states. If the state assembly is non-adiabatic, we imperfectly prepare the target states of our quantum fluid; if the disassembly is non-adiabatic then we imperfectly map them back to the initial qubits. One might expect that such non-adiabaticity would simply reduce the contrast of the resulting manybody Ramsey fringe, but the reality is more subtle: to the extent that the non-adiabaticity is minimal, only a small amount of population is transferred out of the instantaneous eigenstates during assembly, of which some fraction is transferred back during disassembly (see Fig.~\ref{fig:adb}a). This has the effect of adding new frequency components to the Ramsey fringe that provide information about the excitation spectrum of the manybody system.

We investigate this phenomenon in Figure~\ref{fig:adb}b-d by varying the length $\tau$ of our adiabatic assembly and disassembly ramps. In Fig.~\ref{fig:adb}b, we plot the Fourier transform of the Ramsey fringe for the slowest (upper) and fastest (lower) ramps: When the ramp is slow compared with manybody gaps ($\tau=1$\textmu s\,$\gg J^{-1}$), we observe a single frequency component in the Ramsey spectrum indicating preparation of a superposition of only a single pair of states. When the ramp is fast ($\tau=1$ns\,$\ll J^{-1}$), we observe numerous frequency components in the Ramsey spectrum indicating that we have prepared numerous pairs of states that then interfere. In Fig.~\ref{fig:adb}c we plot the Fourier spectrum as we tune the ramp time $\tau$ over three decades, observing the appearance of increasing numbers of peaks as the ramp gets faster. Repeating this experiment with more particles (Figure~\ref{fig:adb}d) reveals fewer total peaks, despite the larger state space accessible with more particles. This occurs because the number of accessible states grows so rapidly that for all but the slowest ramps the features overlap and smear into a continuum.

In practice, achieving the best spectroscopic resolution for the Ramsey signal frequency is a balancing act between: (i) particle loss/dephasing if the protocol takes too long compared to the photon $T_1$/$T_2$ (see SI~\ref{SI:DeviceFabandParams}) and (ii) reduction in the spectral weight of the correct Fourier peak if the adiabatic ramp time $\tau$ is too small and the wrong manybody states are prepared. In order to circumvent decoherence in larger systems, we choose faster ramps that induce some diabatic excitation without obscuring the correct Fourier feature.
\begin{figure*}[ht] 
	\centering
	\includegraphics[width=0.95\textwidth]{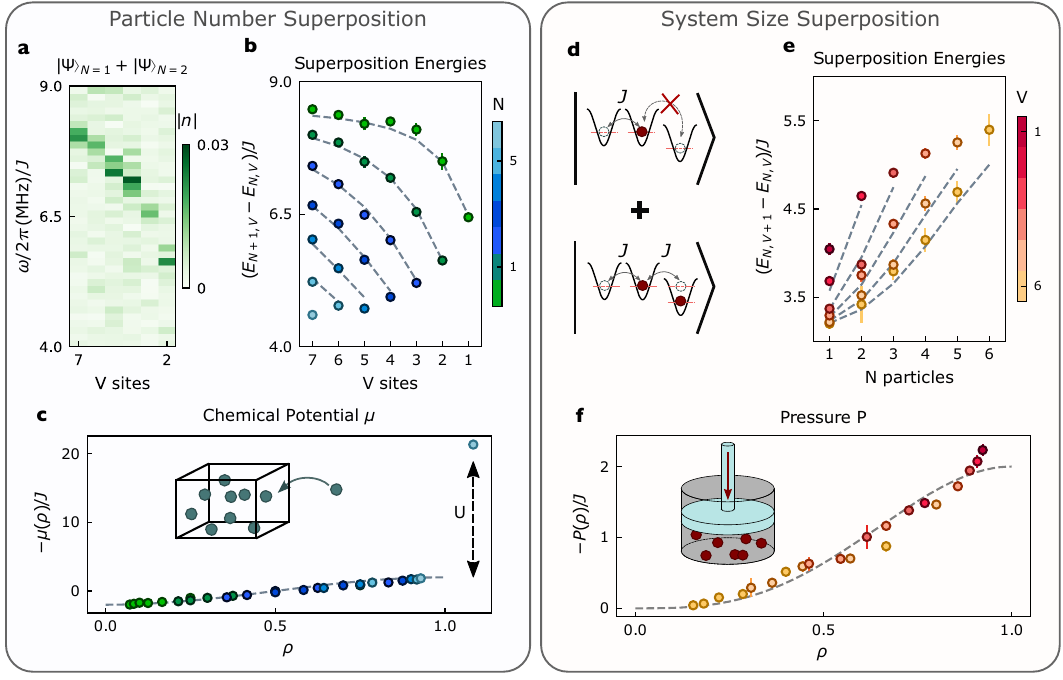}
	\caption{
	\textbf{Spectroscopic Probes of Thermodynamics}. Manybody Ramsey interferometry offers new ways to characterize synthetic quantum matter: \textbf{a...c,} The chemical potential $\mu = E_{N+1, V} - E_{N, V}$ quantifies the energy required to add a particle to a manybody system, and we measure it by interfering states of different particle number. \textbf{a} is a sample dataset showing the Ramsey spectrum for the superposition of $N=1,2$ particles as we vary the total system size $V$. For each $V$, the chemical potential $\mu$ is assigned to the frequency of maximal spectral density, which we plot in \textbf{b} for all fillings up to unit filling $N=0...V-1$, and all system sizes $V=1...7$. In \textbf{c} we replot all data vs the density $\rho\equiv N/V$, finding a collapse onto a universal sinusoidal form (gray) consistent with a free-fermion model~\cite{Rigol2011Bose1DRev}. \textbf{d...f,} The pressure $P = E_{N, V+1} - E_{N, V}$ quantifies the energy required to change the system size, and we measure it by interfering states of different system size. We achieve controlled superpositions of system sizes using the approach shown in \textbf{d}: the controlling site is $U$-detuned such that when it is empty, it is energetically inaccessible, reducing the system size by a single site; when it is filled, it becomes accessible and accordingly increases the system size. Using this site as the control in a manybody Ramsey experiment allows us to extract the energy difference between $N$ particles melted into $V+1$ vs $V$ sites (see SI~\ref{SI:CompressibilityMethods}). Performing this protocol for different volumes $V$ and particle numbers $N$ produces the raw data in \textbf{e}, which we rescale vs density in \textbf{f}, again finding agreement with a free-fermion theory. Error bars, where larger than the data-point, reflect the S.E.M. (see SI~\ref{SI:errorbars}).}
    \label{fig:chem&comp}
\end{figure*}
\section{Extracting Thermodynamic Observables}
\label{sec:thermo}

Having demonstrated the ability to interferometrically extract the energy difference of arbitrarily chosen manybody states, we now apply the technique to the measurement of thermodynamic observables of a quantum fluid. To do this we rely upon the fact that thermodynamic quantities like the chemical potential and the pressure may be understood as the rate of change of the system energy with density, and thus particle number, a quantity which our manybody interferometry technique probes directly.

The chemical potential is the energy required to add a particle to the manybody system at fixed system size $V$, $\mu = E_{N+1, V} - E_{N,V}\approx \left.\frac{\partial E}{\partial N}\right\vert^{V}$: we thus measure $\mu$ by performing manybody Ramsey interferometry between the $N$ and $N+1$ particle ground states. In Fig.~\ref{fig:chem&comp}a we demonstrate this measurement for the superposition of $N=1$ and $N=2$ particles as we vary the number of accessible sites $V$. At each $V$, the Fourier frequency with the largest oscillation amplitude corresponds to the chemical potential. In Fig.~\ref{fig:chem&comp}b we plot this chemical potential vs. $V$, repeating the measurements for particle numbers from an empty system $N=0$ to a filled system $N=V-1$. In Fig.~\ref{fig:chem&comp}c we replot the data versus the density $\rho\equiv\frac{N}{V}$, finding collapse onto a universal (intensive) form $\mu=-2J\cos(\pi \rho)$: adding particles reduces the volume available to each particle, increasing the uncertainty-induced kinetic energy required to add it to the system. These data are consistent with a free-fermion model (see SI~\ref{SI:ThermoAnalytics})~\cite{Rigol2011Bose1DRev, Girardeau1960} modulo small system-size corrections (see SI~\ref{SI:FiniteSize}). Beyond unit filling we observe an additional energy cost $U$ per particle reflecting the incompressibility of the unit-filled Mott-insulating state~\cite{Rigol2011Bose1DRev}.

The pressure is the force required to maintain the fluid at fixed size, or equivalently the energy required to reduce the system size $P = E_{N, V} - E_{N,V+1}\approx-\left.\frac{\partial E}{\partial V}\right\vert^{N}$. To directly measure the pressure we thus need to perform manybody interferometry between systems of different sizes rather than different particle numbers. We achieve this by engineering our ancilla qubit to control the system size: as shown in Figure~\ref{fig:chem&comp}d, the ancilla site is detuned in energy by $U$, ensuring that when it is empty particles cannot tunnel onto it (reducing the system size by 1 site), and when it is occupied particles can tunnel. (Bose-enhanced tunneling onto the occupied ancilla is compensated by Floquet engineering, see SI~\ref{SI:CompressibilityMethods}). The measured pressures for all particle numbers and system sizes are shown in Fig.~\ref{fig:chem&comp}e. They are replotted vs density in Fig.~\ref{fig:chem&comp}, demonstrating that higher densities lead to more uncertainty pressure, again in agreement with a free fermion model anticipated to describe the 1D hardcore bosons in our experiments(See SI~\ref{SI:TGgas}).


\section{Conclusion}
\label{sec:outlook}
We have introduced a probe of synthetic quantum matter that accesses new observables by blurring the boundary between state preparation and measurement. Rather than first preparing a manybody state and then characterizing it, we instead control what manybody state we prepare with an ancilla qubit, coherently \emph{reverse} the preparation procedure to disentangle the ancilla from the manybody system, and sandwich this process within an ancilla Ramsey sequence. This manybody Ramsey interferometry protocol enables direct measurement of energy differences of different eigenstates of the same system, as well as the same eigenstate of different systems. We employ it to directly extract thermodynamic properties of a quantum fluid.

Because the manybody Ramsey protocol relies upon reversible adiabatic assembly of manybody states~\cite{AdbPaper}, it requires only a small factor more coherence time than adiabatic state preparation. In other words, if you can build a state, you can characterize it with manybody Ramsey interferometry.

Marrying these techniques with recent advances in topological quantum matter~\cite{owens2022chiral} will enable probes of fractional statistics~\cite{grusdt2016interferometric}; applying the techniques to glassy~\cite{Dupuis_1DBoseGlassTheory,Meldgin_BoseGlassQuench} or time-crystalline~\cite{zhang2017observation,choi2017observation} phases has the potential to shed light on their structure. We anticipate opportunities to apply manybody Ramsey interferometry to cold atoms, particularly in topological~\cite{leonard2023realization} or fermionic sectors~\cite{chiu2018quantum}. Marrying this tool with a quantum Fourier transform suggests yet more efficient approaches to quantum sensing in manybody systems~\cite{vorobyov2021quantum}. More broadly, this work invites the question: what observables become accessible when \emph{multiple} ancillas are entangled with, and then disentangled from, a quantum material? We envision a future where hitherto unimagined observables are probed by entangling quantum matter with small quantum computers.


\section{Acknowledgments}
This work was supported by ARO MURI Grant W911NF-15-1-0397, AFOSR MURI Grant FA9550-19-1-0399, and by NSF Eager Grant 1926604. Support was also provided by the Chicago MRSEC, which is funded by NSF through Grant DMR-1420709. G.R. and M.G.P acknowledge support from the NSF GRFP. A.V. acknowledges support from the MRSEC-funded Kadanoff-Rice Postdoctoral Research Fellowship. We acknowledge support from the Samsung Advanced Institute of Technology Global Research Partnership. Devices were fabricated in the Pritzker Nanofabrication Facility at the University of Chicago, which receives support from Soft and Hybrid Nanotechnology Experimental (SHyNE) Resource (NSF ECCS-1542205), a node of the National Science Foundation’s National Nanotechnology Coordinated Infrastructure.
We would like to thank Kaden Hazzard for discussions.

\section{Author Contributions}
The experiments were designed by G.R., A.V., J.S., and D.S. The apparatus was built by B.S., A.V., and G.R. The collection of data was handled by G.R. All authors analyzed the data and contributed to the manuscript.

\section{Competing Interests}
The authors declare no competing financial or non-financial interests.





\onecolumngrid

\clearpage
\newpage





\onecolumngrid
\newpage
\section*{Supplementary Information}
\appendix
\renewcommand{\appendixname}{Supplement}
\renewcommand{\theequation}{S\arabic{equation}}
\renewcommand{\thefigure}{S\arabic{figure}}
\renewcommand{\figurename}{Supplemental Information Fig.}
\renewcommand{\tablename}{Table}
\setcounter{figure}{0}
\setcounter{table}{0}


\section{Ramsey Interferometry Measurements}
\label{SI:RamseyMeas}
A Ramsey interferometry experiment on a single qubit measures that qubit's $\omega_{01}$ frequency (relative to the frequency of the tone used to drive the qubit). The experiment sequence is as follows. First, the qubit is driven into an equal superposition of $\ket{0}$ and $\ket{1}$ with a microwave $\pi/2$ pulse. The two states evolve relative to each other for time $T$ with phase $\phi = \omega_{01}T$. A second $\pi/2$ pulse is applied, mapping oscillations around the equator of the Bloch sphere to population oscillations. The qubit population is read out, producing a fringe oscillating at $\omega_{01}T$ minus the frequency of the drive that the qubit is referenced to.

Manybody Ramsey works in a similar fashion. When measuring the energy difference of states with $N$ vs $N-1$ particles when the qubits are all on resonance at the lattice frequency, the energy difference is approximately one lattice photon, $5.2$~GHz. However, the qubit drive to which the energy difference is compared is at the frequency of the ancilla in the staggered position, on the order of hundreds of MHz detuned from the lattice frequency. The time resolution of our AWG is $1$~ns; the maximum frequency we can physically record without aliasing is $500$MHz; the maximum frequency we can comfortably measure is $250$~MHz. In practice, we use a sampling rate of $3-4~\mu$s, which brings our maximum measurable frequency even lower. We thus add a time-dependent virtual phase to our second Ramsey $\pi/2$ pulse in order to virtually change the frequency of the qubit drive used as reference, bringing the measured oscillation frequency below the aliasing limit. 

While measuring states with $V$ vs $V-1$ sites, at first glance it seems like the energy differences involved should be within the same particle manifold. However, because of the way we measure volume superpositions (see SI~\ref{SI:CompressibilityMethods}), the states we compare do differ by a particle, so we measure Ramsey oscillation as above.

To achieve appropriate frequency resolution for experiments involving examining Ramsey Fourier peaks, we record the Ramsey fringe for a minimum of $T=800$~ns in steps of $3-4$~ns for a frequency resolution of at maximum $1.25$~MHz (Fig.~\ref{fig:adb} $2.7$~\textmu s, Fig.~\ref{fig:chem&comp} $804$ns). We choose our time resolution to balance experiment time (long experiments suffer more from frequency drift) and retaining the ability to distinguish frequencies we care about without encountering frequency aliasing.

\section{Thermodynamic Observable Methods}
\label{SI:CompressibilityMethods}
\subsection{Chemical Potential}
\label{SI:CompressibilityMethods:chemPot}
To extract chemical potential, we compare the energy differences of states with $N$ vs $N-1$ particles in a given volume $V$ for a range of volumes and particle numbers by extracting the interference fringe between the relevant states. We compare highest energy eigenstates of each particle manifold, which map to the ground states of a repulsive-$U$ model (see SI~\ref{SI:HighestEState}). Depending on the number of particles, volume, and adiabatic ramp time, this measurement is operated at the edge of the coherence time of our qubits. To extract consistent signal, we play a few tricks. First, which of the qubits participate in the highest energy eigenstate in the disordered configuration depends on which qubits we place highest in frequency. Therefore, we can arbitrarily choose which qubits to use for any given eigenstate (limitations: neighbors need to start properly detuned so that disordered state is separable). For volumes $V$ less than $7$, we can also choose different contiguous sets of qubits as our volume for that experiment. To find the configuration with best coherence and least noise for a given superposition, we cycle through qubits, volume sets (where possible), and ramp times, until we hit a combination that has a peak of prominence $6\sigma$ above background noise after applying a digital low-pass filter cutting off frequency components below $20$~MHz with the Ramsey virtual frequency chosen to fall between $30$~MHz and $100$~MHz (see SI~\ref{SI:RamseyMeas}). We then repeat the experiment to acquire error bars on peaks. 

In order to have small qubit-readout crosstalk, when reading out we place the qubits in a stagger with nearest-neighbor detuning $\gg U$. For Fig.~\ref{fig:PrepSuperpositions} and Fig.~\ref{fig:adb} superpositions were prepared directly from this point. When preparing superpositions of states for Fig.~\ref{fig:chem&comp} however, since some states at higher densities involve nearest neighbors, to avoid unwanted crossings with the $U$-band, qubits were first populated with photons, then rapidly jumped from their readout position to a stagger with nearest-neighbor detuning less than $U$ and qubit position chosen to minimize hybridization. We found we suffered minimally from Landau-Zener transitions (i.e. unwanted population transfer between nearest-neighbors during rapid ramps) and were consistently able to choose staggers where eigenstates still consisted of local qubit states. The adiabatic melt was then performed from this smaller stagger.

Data is plotted with $\omega_{lat}$ subtracted off, such that the theory and data curves are centered around $0$.

\subsection{Pressure}
\label{SI:CompressibilityMethods:pressure}

\begin{figure*} 
	\centering
 	\includegraphics[width=0.5\textwidth]{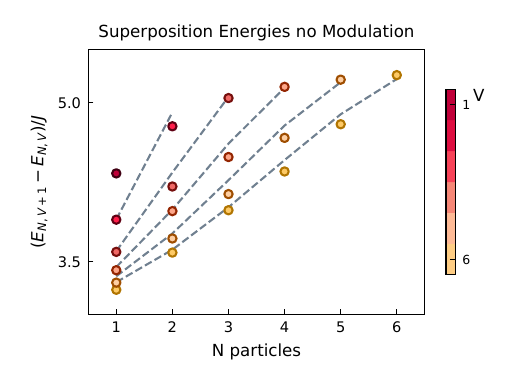}
	\caption{
		\textbf{Interferometry of volume superpositions without modulating to correct for Bose-enhancement}. We measure $E_{N, V+1}-E_{N, V}$ for different particle numbers and volumes without modulating the edge qubit to correct for Bose-enhancement. Our measurements line up well with disorder-free numerics (grey dashed lines), confirming that there is low frequency disorder between the qubits before adding the extra complexity of the modulation step. Error bars, where larger than the data-point, reflect the S.E.M. 
	}
	\label{SI:fig:Vdat_nomod}
\end{figure*} 

We apply a similar procedure to measure pressure, with one extra step. As a reminder, to measure the energy of $N$ particles in $V$ vs $V+1$ sites in order to extract pressure, we detune an edge qubit by the anharmonicity, and place that qubit into a superposition of $\ket{0}$ and $\ket{1}$ to make the qubit $\ket{2}$-state in a superposition of being accessible vs not to the rest of the lattice. However, when tunneling into the $\ket{2}$-state, the tunneling is Bose-enhanced by a factor of $\sqrt{2}$. As a sanity check, we measure energy differences in this configuration and compare to an exact numerical model, where we indeed see the spectrum affected by the Bose-enhanced tunneling to the edge site (see Fig.~\ref{SI:fig:Vdat_nomod}). To suppress this extra tunneling, we frequency modulate~\cite{meinert2016floquet} the edge qubit. When frequency modulating, in the rotating frame, the qubit effectively lives with amplitude given by $J_m(\frac{\epsilon}{2\nu_{sb}})$ at the base frequency plus multiples $m$ of the modulation frequency, where $J$ is a Bessel function of the first order, $\epsilon$ is the strength of modulation in qubit frequency, and $\nu_{sb}$ is the sideband frequency. If one modulates at the right amplitude, one can engineer a $\sqrt{2}$ suppression of tunneling at the base frequency. 

\begin{figure*} 
	\centering
 	\includegraphics[width=0.9\textwidth]{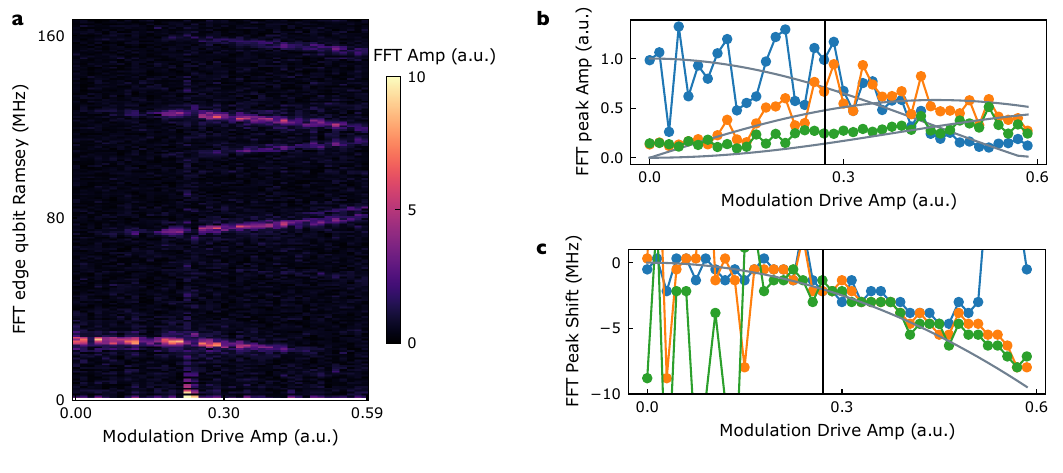}
	\caption{
		\textbf{Modulation Calibrations}. We frequency modulate $Q6$, the lattice edge qubit, at $100$~MHz and measure the corresponding Ramsey signal at different modulation amplitudes. In \textbf{a}, the Fourier transform of the corresponding $Q6$ Ramsey fringes are plotted vs modulation drive amplitude, with a $20$~MHz virtual offset. The peaks correspond to the different sidebands, and the peaks' shift in frequency as a function of amplitude is the DC-offset caused by the the non-linearity in the qubit flux-frequency curve. Several of the peaks at higher frequency are frequency folded/aliased because of finite sampling bandwidth. In \textbf{b}, the first peaks at the first three sidebands are extracted and their amplitude plotted. By fitting these amplitudes to the expected analytic expressions (Bessel functions), we extract the modulation amplitude where the signal at the base qubit frequency (blue) is suppressed by a factor of $\sqrt{2}$, marked by the black vertical line. Using the same fit parameters as for the Bessel functions, we also fit the DC shift for the range of modulation drives in \textbf{c}. This fit is used to extract and correct for the DC shift detuning when modulating by tuning the qubit to compensate.
	}
	\label{SI:fig:modCals}
\end{figure*}

We choose to modulate at $100$~MHz in order to have the higher sidebands be far enough detuned from the lattice to not affect its physics. Because the flux vs current curve of the qubits isn't linear, flux modulating also gives a DC offset term given by $\delta_{DC} = \frac{\epsilon^{2}_{\phi}}{4}\frac{d^{2}\nu_{01}}{d\phi^2}$ where $\epsilon_{\phi}$ is the amplitude of modulation in terms of flux $\phi$ applied to the qubit and $\nu_{01}$ is the frequency of the qubit $\ket{0}$ to $\ket{1}$ transition. We calibrate the flux drive amplitude by fitting single qubit Ramsey frequency components to corresponding Bessel functions and measuring DC offset, see Fig.~\ref{SI:fig:modCals}. 

Applying the modulation from the start of the ramp of the disorder makes adiabaticity quite hard to achieve; the higher sidebands of the modulated qubit can interact with neighbors during the ramp. We found it easier to instead adiabatically ramp qubits onto resonance, and then adiabatically turn on the modulation. In the code loop aiming to find good parameters, we also varied ramp time for turning on modulation.

Similarly to the data set for chemical potential, to get signal for a given superposition, we cycle through qubits, volume sets (where possible), and ramp times, until we hit a combination that has a peak of prominence $6\sigma$ above background noise after applying a digital low-pass filter cutting off frequency components below $20$~MHz with a Ramsey virtual frequency chosen to fall between $30$~MHz and $100$~MHz (see SI~\ref{SI:RamseyMeas}). The exception is for large volume and particle number: for $V=7$ vs $V=6$ sites at filling $N=4, 5, 6$ there are no peaks $6\sigma$ above background noise. We instead choose peaks $4\sigma$ above the background noise and apply a $30$~MHz low pass filter (higher filter since because of our lower peak cutoff we are more susceptible to low frequency noise, and because these energy differences are expected to be higher in frequency).

\subsection{Derivative}
\label{SI:CompressibilityMethods:derivatives}
The $x$-axis point for a measurement comparing $y(x+1)-y(x)$ is $x+0.5$. For example, for $E(N+1, V)-E(N,V)$, the density point in the plot is $\rho=(N+0.5)/V$.


\section{TG Gas and Fermioniziation}
\label{SI:TGgas}
Girardeau showed in 1960 that there is a one-to-one correspondence between impenetrable (i.e. strongly interacting) bosons confined in 1D and non-interacting spinless fermions~\cite{Girardeau1960}. A collection of bosons in the regime where this mapping holds is called a Tonks-Girardeau (TG) gas. 

This mapping between strongly interacting bosons and non-interacting fermions is quite useful, as closed-form expressions for thermodynamic quantities for non-interacting fermions are straightforward to analytically derive. The theory curves presented for pressure and chemical potential in Fig.~\ref{fig:chem&comp} come from the analytic expressions for non-interacting fermions-on-a-lattice. The accuracy of the Girardeau's mapping depends on the relative strength of the interaction term and the tunneling term; terms that break the mapping (to first order they appear like fermion interactions) scale as powers of $J/U$ ~\cite{Cazalilla_Tonks_cnt_lat}. In this experiment $J/U$ = 0.04, and all corrections to energy/chemical potential/pressure are small, on average $1.2$~MHz, the same order of magnitude as deviation from exact numerics ($700-800$~kHz, see SI~\ref{SI:FiniteU}. 

While we use an exact numerical model for the dashed grey theory curves for expected density profiles in Fig.\ref{fig:PrepSuperpositions}, we can also use TG gas formalism to solve for these density profiles as well. We assume a 1D ground state many-body wave function of the Bijl-Jastrow~\cite{bijl1940lowest} form $\Psi_B(\mathbf{x}) = \phi(\mathbf{x}) \varphi(\mathbf{x})$, for $\mathbf{x} = (x_0, x_1,\ldots, x_6)$, where the open boundary condition of our lattice (i.e. the potential well) is captured by the component $\phi(\mathbf{x}) = \prod_{i=0}^{6} \cos(\pi x_i / L)$, and the two-particle component $\varphi(\mathbf{x}) = \prod_{i<j} |x_i - x_j|$ gives the TG gas impenetrable boson requirement~\cite{Rigol2011Bose1DRev}. Using this trial wavefunction, we calculate density profiles for different particle numbers in the potential, and find very close agreement between exact diagonalization of our lattice and the results from the analytic wavefunction~\cite{AdbPaper}. 

\section{Ground vs Highest State}
\label{SI:HighestEState}
We want to explore the properties of the ground states of the repulsive Bose-Hubbard model $\mathbf{H}_\mathrm{BH}(t)/\hbar = J \sum_{ \langle i,j \rangle}{a_i^\dagger a_j } + \frac{U}{2}\sum_i{n_i \left(n_i-1\right)} + \sum_i {(\omega_{lat}+\delta_i(t)) n_i}$. However, the physical Hamiltonian implemented in our experiment has the sign of $U$ and $J$ flipped (ie, we are realizing the attractive Bose-Hubbard model); $H_{\textrm{physical}} = - H_{\textrm{BH}}$. Because the two Hamiltonians differ only by a minus sign, the eigenstates are the same, just with flipped eigenvalues. Since our system is dissipation-less, the dynamics and observables of the highest excited state of our physical model are the same as those of the ground state of the repulsive model (with reversed time and negative values for thermodynamic/energy observables because of the minus sign). Thus, we measure observables of the highest excited state of our system, which maps onto the ground state of a repulsive Bose-Hubbard Hamiltonian. For the rest of the supplement, we refer to the state we measure as the ``ground state"; the analytic values we derive for the ground state thermodynamics quantities match our data up to a minus sign.


\section{Thermodynamic Analytics of 1D Fluids}
\label{SI:ThermoAnalytics}
In this section, we motivate and derive the analytic thermodynamic expressions plotted in grey in Fig.~\ref{fig:chem&comp} assuming our particles act like non-interacting fermions on a lattice, as motivated in SI~\ref{SI:TGgas}. Our data matches these expressions up to a minus sign, as we measure the highest excited state of our system.

In the experiment described in this paper, we compare ground state energies of various particle and volume manifolds. This means we measure ground state observables, i.e. chemical potential, pressure, etc. for $T=0$ and fixed entropy. The following thermodynamic calculations can be done assuming constant temperature and entropy.

The energy eigenvalues of a 1D fermionic lattice with $V$ sites, tunneling $+J$, and open boundary conditions is
\begin{equation}
    E_{k} = 2J\mathrm{cos}\left(\frac{\pi k}{V+1}\right)
\end{equation}
where $k$ are quasi-momenta. The ground state energy of $N$ fermions in a lattice, by the Pauli exclusion principle, is the sum of all single-particle energies from the lowest energy state up, here from $k=N$ to $k=N-V$:
\begin{equation}
\label{Eq:E}
    E_{N} = \sum_{k=N}^{N-V} 2J\mathrm{cos}\left(\frac{\pi k}{V+1}\right) = -J \left[\mathrm{csc}\left(\frac{\sfrac{\pi}{2}}{V+1}\right)\mathrm{sin}\left(\frac{\pi(N+\frac{1}{2})}{V+1}\right) - \frac{1}{2}\right].
\end{equation}
This expression is used to derive thermodynamic quantities below.

To calculate $E$ in the thermodynamic limit, we take $\rho$ constant and send $N$ and $V$ to infinity: 
\begin{equation}
\label{Eq:Eterm1}
    \lim_{N,V\rightarrow\infty}E_{N} =\lim_{N,V\rightarrow\infty} -J \left[\mathrm{csc}\left(\frac{\sfrac{\pi}{2}}{V+1}\right)\mathrm{sin}\left(\frac{\pi(\rho V+\frac{1}{2})}{V+1}\right) - \frac{1}{2}\right] = \frac{-2JV\mathrm{sin}(\pi\rho)}{\pi}.
\end{equation}

The chemical potential at constant entropy is defined as $\mu = \frac{\partial E}{\partial N} \bigg\rvert_{S,V}$. Plugging in the expression for $E$ from~\ref{Eq:E} gives:
\begin{equation}
    \mu = \frac{-J\pi}{V+1}\mathrm{csc}\left(\frac{\sfrac{\pi}{2}}{V+1}\right)\mathrm{cos}\left(\frac{\pi(N+\frac{1}{2})}{V+1}\right)
\end{equation}.

In the thermodynamic limit, sending $N$ and $V$ to infinity while holding $\rho\equiv \frac{N}{V}$ constant, 
\begin{equation}
\label{Eq:Eterm2}
    \lim_{N,V\rightarrow\infty}\mu = -2J\mathrm{cos}(\rho\pi).
\end{equation}

The pressure at constant entropy is defined as $P = -\frac{\partial E}{\partial V} \bigg\rvert_{S,N}$. Plugging in the expression for $E$ from~\ref{Eq:E} yields:
\begin{equation}
    P = \frac{J\frac{\pi}{2}}{(V+1)^2}
    \mathrm{csc}\left(\frac{\frac{\pi}{2}}{V+1}\right) \left[\mathrm{cot}\left(\frac{\sfrac{\pi}{2}}{V+1}\right)
    \mathrm{sin}\left(\frac{\pi(N+\frac{1}{2})}{V+1}\right) 
    - (2N+1)\mathrm{cos}\left(\frac{\pi(N+\frac{1}{2})}{V+1}\right)
    \right]
\end{equation}.

In the thermodynamic limit, holding $\rho$ constant while sending $N$ and $V$ to infinity, 
\begin{equation}
\label{Eq:Eterm3}
    \lim_{N,V\rightarrow\infty}P = 2J\left(\frac{\mathrm{sin}(\pi\rho)}{\pi}-\rho\mathrm{cos}(\rho\pi)\right).
\end{equation}

\section{Souces of Deviation from Non-Interacting Fermion Analytics}
\subsection{Finite N and V effects}
\label{SI:FiniteSize}
\begin{figure*} 
	\centering
 	\includegraphics[width=0.9\textwidth]{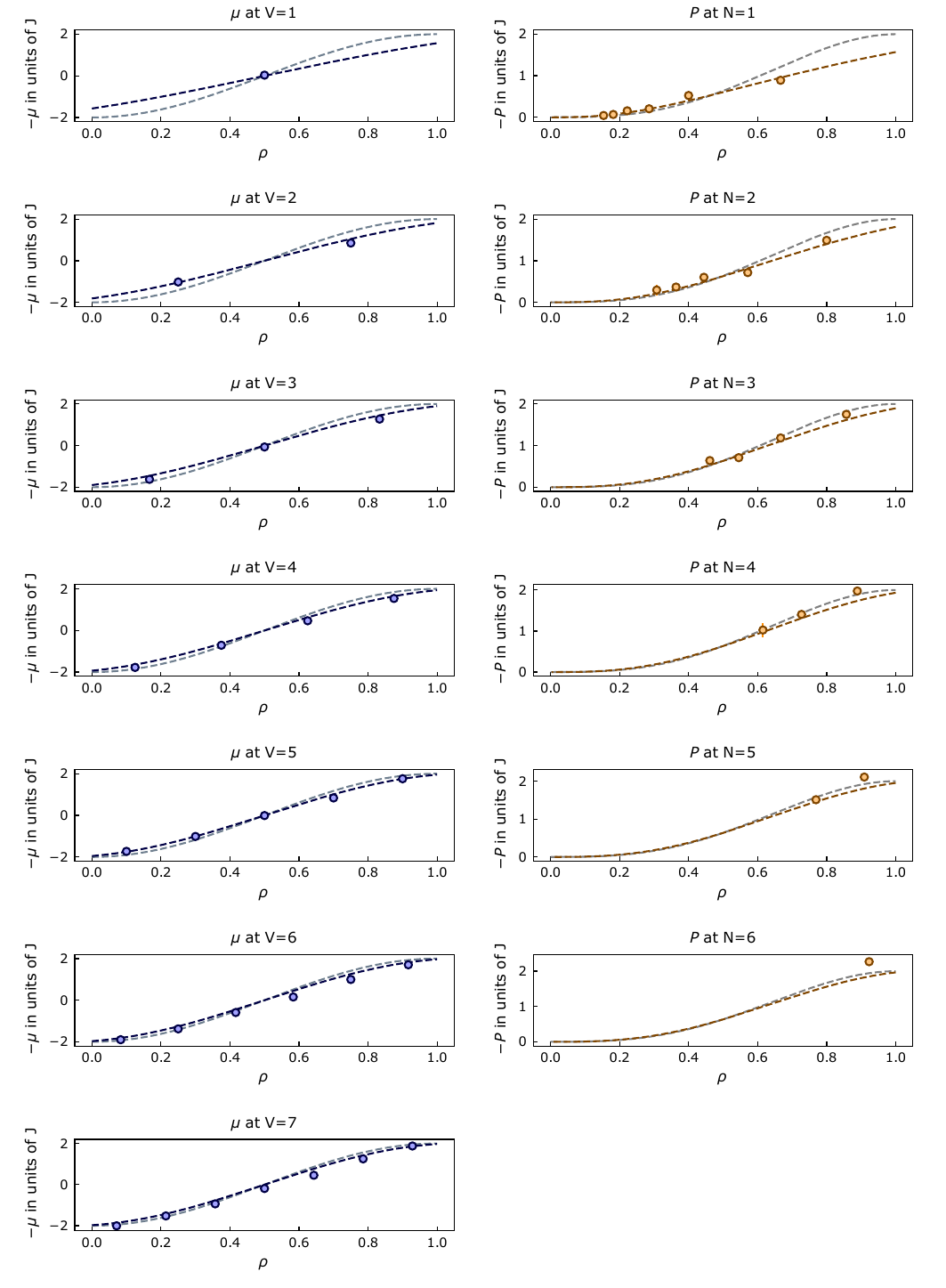}
	\caption{
		\textbf{Finite Size N and V Effects in Chemical Potential and Pressure}. Grey dashed lines correspond to free fermion analytics in the thermodynamic limit, dashed colored lines correspond to free fermion analytics in finite size limit indicated by subplot title, and circles correspond to data. Error bars, where larger than the data-point, reflect the S.E.M.
	}
	\label{SI:fig:finiteSize}
\end{figure*}

The gray theory curves plotted in Fig~\ref{fig:chem&comp} are theory for thermodynamic limit. However, our system is finite in size, and so deviates from this limit. Our data is better captured by finite-size analytics, see Fig~\ref{SI:fig:finiteSize}. 

\subsection{Finite U/J Effects in a Finite Size System}
\label{SI:FiniteU}
\begin{figure*} 
	\centering
 	\includegraphics[width=0.95\textwidth]{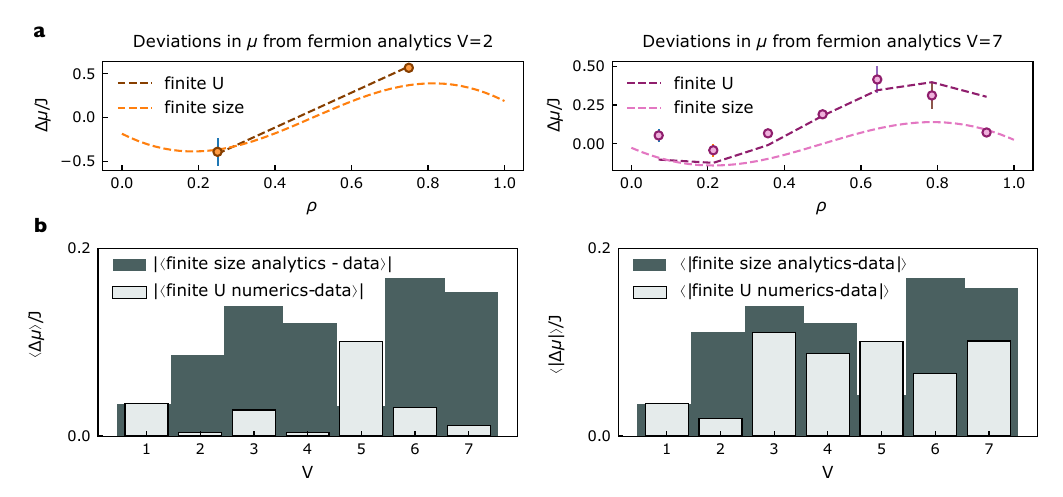}
	\caption{
		\textbf{Example of finite $U/J$ effects in chemical potential}. In \textbf{a}, we plot the deviation from the infinite size fermion chemical potential for finite size analytics (dashed light line), finite size \& finite $U$ numerics (dashed dark line), and our data (circles). In both representative volumes $V=2$ and $V=7$, it emerges that while there is scatter in the data, data agrees more closely with finite $U$ \& finite size numerics than just the finite size theory. In \textbf{b}, in the first panel, for each volume, we measure the average deviation between data and finite size fermion (ie infinite $U$) analytics (dark grey) and between data and numerics calculated with our finite experimental value of $U$ (light grey). In the second panel, we plot the average of the absolute value of the deviation. These panels illustrate that while the magnitude of the deviation in the data between finite $U$ numerics and infinite U analytics is similar, the data still agrees more closely with finite U numerics. Error bars, where larger than the data-point, reflect the S.E.M.
	}
	\label{SI:fig:finiteU}
\end{figure*}

Zooming in further reveals that at the next level of correction, our data also deviate from even the finite-size correction because of the interaction energy U being finite; the mapping to non-interacting fermions is not perfect, see Fig.~\ref{SI:fig:finiteU}. 


\section{Anharmonicity Disorder}
\label{SI:UbandDisorder}
We measured the Mott insulator gap at all lattice volumes, with $Q1$ as the qubit in a superposition of $\ket{1}$ and $\ket{2}$. However, because of strong variation in the qubits' anharmonicities, lattice configurations beyond $V=2$ that include $Q2$ were effectively restricted to $Q0$ and $Q1$: $Q2$'s $\ket{2}$ is $26$~MHz detuned from its neighbors, causing data at higher particle number to deviate from values expected in an uniform-$U$ case. 
\begin{center}
{\renewcommand{\arraystretch}{1.2}
\begin{table}
\begin{tabular}{|c|c|c|c|c|c|c|c|}

\hline
 \textbf{Qubit} & \textbf{1} & \textbf{2} & \textbf{3} & \textbf{4} & \textbf{5} & \textbf{6} & \textbf{7} \\
\hline
 $U_{\textrm{lattice}}/2\pi$ (MHz) & -236 & -235 & -209 & -234 & -236 & -231 & -225  \\ 
\hline
 $(U_{\textrm{lattice}}-U_{Q1})/2\pi$ (MHz) & -1 & 0 & 27 & 1 & -1 & 4 & 10  \\ 
\hline
\end{tabular}
\caption{
	\textbf{Disorder in Qubit Anharmonicity}
}
\label{SI:UbandDis}
\end{table}
}
\end{center}

\begin{figure*}
	\centering
 	\includegraphics[width=0.55\textwidth]{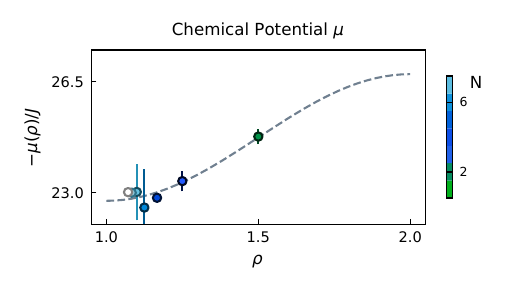}
	\caption{
		\textbf{Chemical Potential Above Mott Insulator Gap}. We plot the energy difference between the lattice being fully filled and the lattice being fully filled with one particle band of second excited states $U$ below the first band. Because of Bose enhancement, the effective tunneling in this second band is $2J$, reflected in the grey dashed lines from non-interacting fermion analytics. Because of lattice site disorder, data deviates from free fermion analytics at higher volumes (i.e. lower densities in plot above since particle number here is fixed at $N=1$). Error bars, where larger than the data-point, reflect the S.E.M.
	}
	\label{SI:fig:Uband}
\end{figure*} 

\section{Same Particle Manifold Superpositions}
\label{SI:SameManifoldSup}

\begin{figure*} 
	\centering
 	\includegraphics[width=0.95\textwidth]{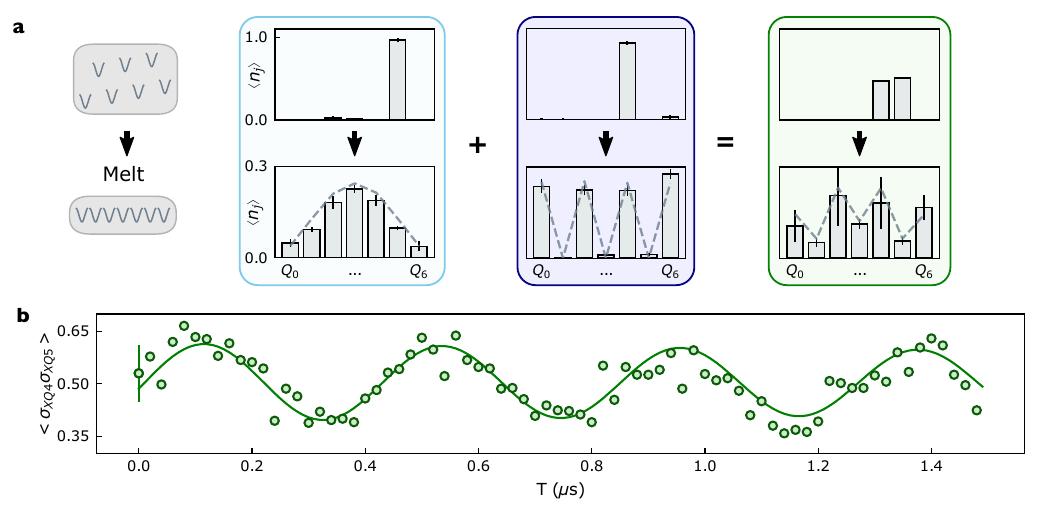}
	\caption{
		\textbf{Same Particle Manifold Superpositions}. In \textbf{a}, we create and measure density profiles for two different one-particle states, and their superposition, at the disordered stagger position and at the lattice position. We then record the Ramsey beating between the two states by measuring $\langle \sigma_{XQ4} \sigma_{XQ5}\rangle$ in \textbf{b}. Representative error bars reflect the S.E.M.
	}
	\label{SI:fig:SamePart}
\end{figure*} 

We perform a two-qubit gate, preparing the state $\frac{1}{\sqrt{2}} (\ket{01} + \ket{10})$ between two qubits, to create a superposition of these eigenstates at when the qubits are far detuned from each other, in the stagger configuration. There are several ways to enact a two qubit gate: the way we choose here is to $\pi$-pulse one qubit, bring it in resonance with its neighbor for half the $J$ tunneling time, and then jump both qubits back to the stagger. We then proceed with the reversible adiabatic ramp protocol as normal. In Fig.~\ref{SI:fig:SamePart}\textbf{a} we prepare and measure density profiles for two different single particle states at the stagger and at the lattice degeneracy point where all the qubits are on resonance, as well as the density profile for their superposition. We extract the Ramsey trace by measuring $\langle \sigma_{XQ4} \sigma_{XQ5}\rangle$ in Fig.~\ref{SI:fig:SamePart}\textbf{b}, which is within $3$~MHz of the expected value.

\section{Compressibility}
\label{SI:Compressibility Data}

\begin{figure*} 
	\centering
 	\includegraphics[width=0.55\textwidth]{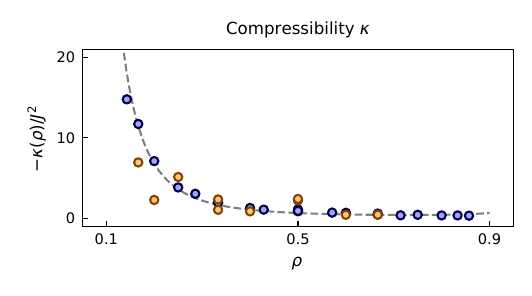}
	\caption{
		\textbf{Compressibility}. We plot the quantum fluid compressibility at different density points by differentiating pressure data (orange) and chemical potential data (blue). Error bars, where larger than the data-point, reflect the S.E.M.
	}
	\label{SI:fig:comp}
\end{figure*} 

The compressibility reflects how much the pressure changes with volume. In Fig.~\ref{SI:fig:comp} we compare compressibility computed by differentiating pressure data ($\kappa_s^{-1}=-V\frac{\partial P}{\partial V}$) to compressibility computed by differentiating chemical potential data ($\kappa_s^{-1} = V \rho^2 \frac{\partial \mu}{\partial N} \rvert_{S}$). The agreement that we find further validates that the equation of state is intensive (dependent on particle number and system size only through density).

We calculate the compressibility by taking numeric $N$ and $V$ derivatives of $\mu$ and $P$. The thermodynamic compressibility at constant entropy and particle number is defined as 
\begin{equation}
    \kappa_s^{-1} = - V \frac{\partial P}{\partial V} \bigg\rvert_{S, N}.
\end{equation}

However, with this expression, we can only take a numeric derivative of our pressure data set with respect to volume; it would be be nice to use our chemical potential data set as well.

In the thermodynamic limit, the Gibbs-Duhem expression holds, which states
\begin{equation}
    0 = - V\partial P + S\partial T + N\partial \mu.
\end{equation}
Since we are at $T=0$, this implies $\partial P =N/V \partial\mu = \rho\partial\mu$. Therefore, 
\begin{equation}
    \kappa_s^{-1} = - V \frac{\partial P}{\partial V} \bigg\rvert_{S, N}
    = - N \frac{\partial \mu}{\partial V} \bigg\rvert_{S, N}
    = - N \frac{\partial}{\partial V} \bigg\rvert_{S, N}\frac{\partial E}{\partial N} \bigg\rvert_{S, V}.
\end{equation}

We flip the order of the derivatives, and once again replace $\partial P =N/V \partial\mu = \rho\partial\mu$ to get the expression $\kappa_s^{-1} = - V \rho^2 \frac{\partial ^2 E}{\partial N^2} \bigg\rvert_{S, V}$ ~\cite{Cazalilla_1Dgas}.

\section{Experiment Details}
\label{SI:DeviceFabandParams}
\begin{center}
{\renewcommand{\arraystretch}{1.2}
\begin{table}
\begin{tabular}{|c|c|c|c|c|c|c|c|}
\hline
 \textbf{Qubit} & \textbf{1} & \textbf{2} & \textbf{3} & \textbf{4} & \textbf{5} & \textbf{6} & \textbf{7} \\
\hline
 $U_{\textrm{lattice}}/2\pi$ (MHz) & -236 & -235 & -209 & -234 & -236 & -231 & -225  \\ 
 \hline
 $J_{i, i+1}/2\pi$ (MHz) & -9.62 & -9.58 & -9.63 & -9.74 & -9.76 & -9.63 & --  \\ 
 \hline
 $T1 (\mu s)$ & 14.6 & 35.5 & 57.7 & 28.4 & 60.3 & 54.7 & 40.0  \\
 \hline
 $T2^* (\mu s)$ & 0.85 & 0.64 & 1.31 & 0.77 & 3.57 & 0.84 & 1.4  \\
\hline
\end{tabular}
\caption{
	\textbf{System Parameters}
}
\label{SI:systemparameters}
\end{table}
}
\end{center}

The optimal lattice frequency varied on a scale of weeks depending on the frequency distribution of low lifetime defects. For data taken for Fig.~\ref{fig:PrepSuperpositions} we used lattice frequency $5.31$~GHz, for Fig.~\ref{fig:adb} we used lattice frequency $4.820$~GHz, and for Fig.~\ref{fig:chem&comp} we used lattice frequency $5.0$~GHz.
Our qubit $T1$s and $T2$s similarly varied over time, with the average $T1=40$~\textmu s and average $T2^{*}= 1.3(\mu s)$. 
Our qubit T2 times when all the qubits are on resonance improve, since because of the avoided crossings the eigenvalue vs flux curves become flatter (we generate our own sweet spot). We did not quantitatively measure this effect.

See the supplement of our previous work~\cite{AdbPaper} for further details on device fabrication and parameters, DC \& RF flux calibrations, flux line transfer function correction, readout methods, and cryogenic + room temperature wiring. The RF crosstalk measured at $100$~MHz for experiments involving modulation in this work is lower, close to $2-3$\%. 


\section{Disorder Correction}
\label{SI:Disorder Correction}
Ensuring that the qubits in the on-resonance lattice configuration are degenerate in frequency is extremely important when measuring pressure and chemical potential. For several densities, the energy differences between ground states are on the order of a few MHz; lattice disorder can cause significant error in the quantities we are attempting to measure. Using our RF flux crosstalk matrix correction and measured qubit $\omega_{01}(\phi)$ relations, we are able to place qubits within $1-2$~MHz of the desired lattice frequency.  To ensure we hit lattice degeneracy to the required precision, we feed back on the local $\omega_{01}(\phi)$ relations by comparing our manybody profiles to expected theory. To ensure our corrections are robust, we feed back on a full set of manybody profiles at $V=7$: the ground state for $N=1$, $N=2$, and $N=3$. Using this method, we are able to achieve error on the order of $200-400$~kHz, which is the same order of magnitude as experiment-to-experiment qubit frequency drift.

It is also important that the qubits not only be on resonance with each other, but also that we know what lattice frequency they are being placed at after the round of corrections described above. Energy differences between eigenstates of different particle number depend on what energy the particles are at. To ensure we are normalizing correctly (see SI~\ref{SI:ThermoAnalytics}), it is important that we actually be placing our lattice at the expected frequency. Feeding back on profiles helps correct for relative detuning between qubits, but does not give us insight into the absolute lattice frequency. To measure this quantity, we first measure via standard Ramsey interferometry the frequency of individual qubits brought to the lattice one at a time using the disorder corrections calculated from feedback back on profiles. Usually there is some scatter (from an imperfect crosstalk matrix); we take the average of this scatter. Recent results have shown that using machine learning on flux crosstalk matrices allows for very low disorder~\cite{CoraFlux}, this would be a better solution going forward.

\section{Error and Uncertainty Calculations}
\label{SI:errorbars}
For each experiment in Fig.~\ref{fig:PrepSuperpositions}, we measure 2000 shots, bin the shots, apply relevant confusion matrices, and extract the averaged quantity of interest (in this case, Ramsey traces). We then repeat the experiment 10-11 times. Because of the long ramp times and wait times in the experiment, very small frequency variations experiment-to-experiment cause large phase variations in our measurement. We calculate the starting phase in each trace from fits, and numerically zero the phase.  We then calculate the mean of the averages and standard deviation of the resulting traces (i.e. calculate the standard error of the mean, or S.E.M.) for the values and error bars that we report in this figure. The experiment repetitions are performed close in time (typically within a 10-30 minute span) so that our error bars are not affected by slow experimental drifts over hours or days.

For error bars in Fig.~\ref{fig:chem&comp}\textbf{b} and \textbf{e} come from repeating the procedure described in SI~\ref{SI:CompressibilityMethods} 3 times and 10-15 times respectively, then taking the mean and standard deviation of the collections of measured peaks. We then propagate the error to gain the error bars in Fig.~\ref{fig:PrepSuperpositions}\textbf{c}, \textbf{f}, and \textbf{g}. The exception is the $U$-detuned point in \textbf{c}, where we did not have enough repetitions of the data point. There, the error bar is the variance in a Lorentzian fit to the measured Ramsey peak.

Note that our error bars in Fig.~\ref{fig:chem&comp}\textbf{b} and \textbf{e} are smaller than points' deviation from exact numerical models of our system (average deviation of $700$~kHz for particle number superpositions and $800$~kHz for volume superpositions). This is likely because of residual frequency disorder (see SI~\ref{SI:Disorder Correction}).

\twocolumngrid

\clearpage
\bibliographystyle{naturemag} 
\bibliography{references}

\subsection{Data Availability}
The experimental data presented in this manuscript are available from the corresponding author upon request, due to the proprietary file formats employed in the data collection process.
\subsection{Code Availability}
The source code for simulations throughout are available from the corresponding author upon request. 
\subsection{Additional Information}
Correspondence and requests for materials should be addressed to D.S. (dschus@stanford.edu).

\
\end{document}